\documentclass[english,twocolumn]{revtex4}
\usepackage[T1]{fontenc}
\usepackage[latin1]{inputenc}
\usepackage{graphicx}
\usepackage{amssymb}
\usepackage{babel}
\makeatother
\begin{document}

\title{Edge-functionalized and substitutional doped graphene nanoribbons: electronic and spin properties}
\author{F. Cervantes-Sodi, G. Csányi, S. Piscanec, A. C. Ferrari}
\affiliation{Department of Engineering, University of Cambridge,
Cambridge, UK }

\begin{abstract}

Graphene nanoribbons are the counterpart of carbon nanotubes in
graphene-based nanoelectronics. We investigate the electronic
properties of chemically modified ribbons by means of
density functional theory. We observe that chemical modifications
of zigzag ribbons can break the spin degeneracy. This promotes the onset
of a semiconducting-metal transition, or of an half-semiconducting state, with the two spin channels having a different bandgap, or of a spin-polarized half-semiconducting state---where the spins in the
valence and conduction bands are oppositely polarized. Edge
functionalization of armchair ribbons gives electronic states a
few eV away from the Fermi level, and does not significantly
affect their bandgap.
N and B produce different effects, depending
on the position of the substitutional site. In particular, edge
substitutions at low density do not significantly alter the
bandgap, while bulk substitution promotes the onset of
semiconducting-metal transitions.  Pyridine-like defects induce a
semiconducting-metal transition.

\end{abstract}

\maketitle

\section{Introduction }
Graphene is the latest carbon allotrope to be
discovered~\cite{Nov438(2005),Nov315(2007),Nov306(2004),GeimRevNM,
CastroNetoRev}, and it is now at the center of a significant
experimental and theoretical research effort. In particular,
near--ballistic transport at room temperature and high carrier
mobilities (between 3000 and 100000 cm$^{2}$/Vs)
~\cite{Nov438(2005),Zhang438(2005),MorozovNov(2007)} make it a
potential material for nanoelectronics~\cite{Lemme, Zhang86, Han,
Chen}, especially for high frequency applications.

It is now possible to produce graphene samples with areas exceeding
thousands of square microns by means of micro-mechanical cleavage of
graphite, and much larger by "expitaxial" growth on SiC,
~\cite{Land264,Entani88,Nov306(2004),Berger2004,Berger2006,Ohta,Nagashima291}.
An ongoing effort is being devoted to large scale production and
growth on different substrates of choice~\cite{Rosei,Marchini,ZhouLanzara,Vazquez,OuyangGuo}.

Graphene can be readily identified in terms of number and
orientation of the layers by means of elastic and
inelastic light scattering, such as Raman
~\cite{ACFRaman,Pisana,Casiraghi_condmat_2007,PiscanecPRL,Latil}
and Rayleigh spectroscopies ~\cite{CasiraghiNL,GeimAPL}.  Raman
spectroscopy also allows monitoring of doping and
defects~\cite{ACFRamanSSC,DasCM,Pisana,Casiraghi_condmat_2007}.
Once identified, graphene layers can be processed into nanoribbons
by lithography ~\cite{Han,Berger2006,Lu75,Lemme}.

As for carbon nanotubes (CNTs)~\cite{Saito, Reich}, electron confinement
modifies the electronic structure of graphene, when this is cut
into nanoribbons (GNRs) \cite{Nakada1996, Fujita1996, Miyamoto59,
Wakabayashi59, Nakada1998, SonPRL, Pisani}.  When GNRs are
cut from a single graphene layer, their edges could in general
consist of a combination of regions having an armchair or a zigzag
geometry~\cite{Niimi,Kobayashi2005,SolsGuineaCastroNeto,NiimiASS}.
If a ribbon is uniquely limited by one of these edges, it is
defined either as an armchair GNR (AGNR) or as a zigzag GNR (ZGNR)
(See, \emph{e.g.}, Fig.~\ref{Fig 1})~\cite{Nakada1996, Fujita1996,
 NiimiASS}.

GNRs are the counterpart of nantubes in graphene nanoelectronics.
Indeed, the confinement of the electronic wavefunctions and the
presence of the edges open a bandgap, making them suitable for the
realization of devices. Due to their potential technological
applications, their electronic structure has been widely
investigated~\cite{Fujita1996, Miyamoto59, Wakabayashi59,
Sasaki88, Fujita1997, Nakada1996, Nakada1998, SonPRL, Lee, Pisani,
Okada2001, Sonnat, Yang0706, Prezzi, Hod, Kan, Rudberg,
Hodnanodots, Martin, Areshkin, Gunycke, Barone, Martins,
HuangDuan, Zheng75}, with particular attention to the
factors determining the presence and the size of the gap.

CNTs can be metallic or
semiconducting, depending on their chirality. This could lead to a fully carbon based electronics, where
semiconducting tubes are used as channels and
metallic ones as interconnects \cite{AvourisII,Graham, Avouris}.
However, the present lack of control on the chirality prevents
to engineer their electronic properties on-demand, and is a major
barrier to industrial implementation.

GNRs could combine the exceptional properties
of graphene, and the possibility of being
manufactured by means of industry-amenable large scale top down
planar technologies, which would uniquely define their chirality.

Introducing impurities and functional groups can be
an effective way to control electronic properties of GNRs. On the other hand, covalently bonded impurities
are likely to be a by-product of the production
processes.

Here, we investigate the effect of chemical disorder on the band
structure of ZGNRs and AGNRs. We first study edge
functionalization of ZGNRs with a set of different radicals,
elucidating the role played (i) by the ribbons width, (ii) by the
concentration of the functional groups along the ribbon's edges,
and (iii) by the presence of one- and two-sided edge
functionalization.
We find that edge functionalization breaks the spin degeneracy, leading to four
possible outcomes.
In the first case, the ribbon maintains its semiconducting nature,
and the top of the valence band and the bottom of the conduction
band belong to the same spin channel. We refer to this as \emph{``spin-selective half-semiconductivity''}~\cite{Rudberg}.
In the second case, the ribbon is still semiconducting, but the top
of the valence band and the bottom of the conduction band belong
to opposite spin channels. We refer to this as
as \emph{``spin-polarized half-semiconductivity''}~\cite{Prigodin}.
In the third and the fourth cases, the band gap of one or both spin
channels closes. In the latter case, the ribbon undergoes a
\emph{semiconductor-metal transition}, while in the former, it
behaves as a \emph{half-metal}~\cite{Groot,Sonnat}.
Edge functionalization on AGNRs at the moderately low densities
we studied has negligible effect on the gap.

We then consider chemical doping on H-terminated AGNR and ZGNR.
AGNRs are always semiconducting, so, in principle, B, N, or O atomic
substitutions can result in some form of electrical doping.
However, given the particular nature of the ribbons, it is not
clear \emph{a-priori} whether chemical doping of AGNRs would
result in the formation of defect states within the gap, like in
carbon nanotubes~\cite{Carrol1998,Yi,Nevidomskyy2003,Czerw} and
bulk semiconductors, or if it would rather cause a shift of the
Fermi level similarly to what observed in
graphene~\cite{Casiraghi_condmat_2007, Tan_Kim_condmat_2007, MorpurgoPC}.
Thus, we investigate in detail the effects of atomic substitutions
on the electronic structure of AGNRs. We find that upon B and N
edge-substitution, impurity levels appear far away from the band
gap, slightly renormalizing its width. On the contrary, N and B
substitution in the bulk of the ribbons result in impurity levels
near the gap, and in a shift of the Fermi energy within one of the
continuum levels of the pristine ribbon. Atomic substitution in
edge functionalized ZGNR and pyridine-like impurities in AGNR
induce semiconductor-metal transitions for high impurity
densities, but do not give impurity levels close to the top or
bottom of the gap, unlike the case of nanotube doping.

\section{Background}

\subsection{Band structure of GNRs}

GNRs can be thought as single wall CNTs cut along a line parallel to their
axis and then unfolded into a planar geometry. Since on a graphene
plane the zigzag and armchair directions are orthogonal, this
procedure transforms an armchair CNT into a ZGNR and a zigzag
CNT into an AGNR.
However, this correspondence between the geometry of
CNTs and GNRs is not entirely reflected in their electronic properties.
Indeed, in contrast to CNTs, where armchair tubes are always
metallic and zigzag could be metallic or semiconducting depending
on the chiral angle~\cite{SaitoAPL}, the earliest theoretical
studies of the electronic structure predicted all ZGNR to be metallic, while AGNR
were expected to be
semiconducting~\cite{Fujita1996,Miyamoto59,Wakabayashi59,Sasaki88,Fujita1997,Nakada1996}.
The metallic character of ZGNR was attributed to the presence of a
high density of edge states at the Fermi
energy~\cite{Fujita1996,Sasaki88,Miyamoto59,Wakabayashi59}.

More recently, spin-polarized DFT calculations have found that the
ground state of ZGNR has a anti-ferromagnetic
configuration (AF), where electronic states with opposite spin are
highly localized at the two GNR
edges~\cite{SonPRL,Lee,Pisani,Fujita1996,Okada2001} and are
responsible for the opening of a gap.
It was shown that the magnetic instability energy of ZGNRs, i. e.
the energy difference between the ferromagnetic and the non
magnetic ground state, increases from $\sim$ 0.25 eV/unit cell up to
$\sim$ 0.37 eV/unit cell when $N$ increases from 5 to 30, and then
stabilizes~\cite{Pisani}. The energy
difference between the ferromagnetic and the AF
ground state, however, are much smaller, with the AF state being
more stable than the ferromagnetic by 4.0, 1.8, 0.4meV/unit
cell when $N$ is 8, 16, 32~\cite{SonPRL}.
Since $k_BT$ at room temperature corresponds to $\sim$25 meV, this
indicates that ZGNRs at room temperature are stable in a magnetic state. As
temperature decreases the anti-ferromagnetic state becomes favored over the
ferromagnetic one. Thus DFT predicts both AGNRs
and ZGNRs to always have a direct nonzero band
gap at least at low temperatures~\cite{SonPRL,Pisani,Sonnat}.
In CNTs the gap scales
inversely with diameter~\cite{OdomLieber}. Similarly,
in GNRs the gap scales inversely with the width~\cite{Nakada1998, Zheng75, Barone, SonPRL, Pisani},
closing, as expected, for infinite graphene.

The electronic structure of GNRs was also studied using
GW calculations~\cite{Yang0706,Prezzi}.  As for CNTs \cite{SpataruPRL}, the better description of the
electron-electron interaction provided by GW
increases the computed GNRs electronic gap. Furthermore,
calculations based on the Bethe-Salpeter equation proved that, due
to the presence of excitons, in GNRs the optical gap is
significantly smaller than the single particle
one~\cite{Yang0706,Prezzi}.
However, apart from the gap renormalization, all the main results
obtained within DFT, including the AF ground-state of ZGNRs and
the width dependence of the bandgap, are confirmed by the
computationally more expensive GW~\cite{Yang0706,Prezzi}.

The presence of a gap inversely proportional to the ribbon's width
was experimentally observed, for ribbons in the 10 to 500nm range,
by means of temperature dependent conductance (G)
measurements~\cite{Han,Chen,Ozyilmaz_Kim_condmat_2007}. The gaps $(E_g)$ measured in these experiments are larger than those
predicted by theory, and their width scales as
$E_g$=$A$($W$-$W*$)$^{-1}$, with $W$ being the ribbon's width and $A$ and
$W$* fit constants~\cite{Han}. In the same samples, the
conductivity $G$ was found to scale as $G$=$B$($W$-$W_0$), with
$B$ and $W_0$ fit constants. For T=1.6K, it was found that
$W$*$\sim$$W_0$=16nm~\cite{Han}.
This suggests that structural disorder at the edges caused by etching, or inaccuracies in the width determination due to
over-etching underneath the mask, can result in an
``effective transport width'' smaller than the ribbon's nominal
width. Within this picture, $W_0$ and $W$* can be considered
as the reduction of the effective width with respect to
the nominal width~\cite{Han}.
Based on these results, some authors have also proposed that the size
of the GNRs bandgap can be explained not only by
confinement, but also by Coulomb blockade originated by the roughness at the edges of the
ribbons~\cite{SolsGuineaCastroNeto}. This would also determine the value of
$W$*$\sim$$W_0$~\cite{SolsGuineaCastroNeto}.
Furthermore, it was also found that the minimum
conductivity as a function of the gate changes with the environment or the GNR thermal history~\cite{Chen}.

Finally, the electronic structure of mono-atomic step edges of
graphite was also probed by scanning tunnelling microscopy (STM)
and spectroscopy (STS).
This showed a large electronic density of states (DOS) at the
Fermi energy $(E_F)$ of zigzag steps, while a bangap was observed
on armchair edges~\cite{Niimi,NiimiASS,Kobayashi2005,Giunta114}.
These results seem to corroborate the simple tight-binding models
predicting an accumulation of electronic states at $E_F$ in
ZGNRs~\cite{Fujita1996, Sasaki88,Niimi,Kobayashi2005}, and to
contradict the presence of the AF bandgap predicted by DFT and GW.
However, it is important to recall that such experiments were
performed on monoatomic steps of bulk graphite, which correspond to a
semi-infinite ribbon. Since the onset of an antiferromagnetic
state requires the presence of wave-function with opposite spins
localized on the two opposite edges of a ribbon, such state cannot
be present in a semi-infinite system with just a single edge.

\subsection{Spin properties}

Ref.~\cite{Sonnat} showed, using DFT, that applying an electric field
parallel to the ribbon's plane results in an opposite local
gating of the spin states on the two edges of the ZGNR. The required
critical field was estimated to be 3.0V/$W$, $W$ in \AA~\cite{Sonnat}.
Indeed, the in-plane field lifts the spin degeneracy by reducing
the band gap for one spin channel, while the other experiences
a gap widening. This promotes the onset of half-metallicity~\cite{Groot,Park}
or induces a spin-selective semiconducting behavior, where the band gap of one
spin is bigger than the other. Which of these states is
established depends on the intensity of the applied field~\cite{Hod,Sonnat,Kan} and on the ribbon's length
~\cite{Rudberg,Kan,Hodnanodots}. These phenomena may have
important applications for the fabrication of spin filtering
devices\cite{Wolf2001}.

Half-metals are materials in
which the current can be completely spin polarized as a
result of the coexistence of a metallic nature for electrons with
one spin orientation and an insulating nature for electrons with the
other spin~\cite{Groot,Sonnat}. Some ferromagnetic metals were predicted to have this
behavior~\cite{Groot} and were first observed experimentally in a
manganese perovskite~\cite{Park}.
Similarly, there can be half semiconducting states in which the electron spins in the
valence and conduction bands are oppositely polarized~\cite{Prigodin}.
Since the development of efficient devices for spin injection,
transport, manipulation and detection is a fundamental requirement
for the incorporation of spin into existing semiconductor
technology and to achieve spin based quantum computing and
spintronics~\cite{Wolf2001}, substantial efforts were made
in to find half-metallic materials~\cite{Wolf2001,Fang}. It is thus of great interest to identify such states in GNRs.

\subsection{Chemical modifications}

Chemical functionalization and substitutional doping have been
investigated for many years in nanotubes~\cite{HirschFunc,
Georgakilas, Carrol1998,Yi,Nevidomskyy2003,Czerw,LeeMarzari}, with the aim of
tailoring their properties for sensing, transport, chemical and optical
applications or to incorporate them into polymers
~\cite{Kong_sensing,Dai_transport,Star_polymer,Scardaci,HasanJPCC}.
It is thus natural to do a similar investigation for GNRs.

Various theoretical studies of the influence of edge
geometry and chemical modifications on the transport properties of
ribbon junctions have been reported so far~\cite{Xu,Wakabayashi,Monoz,Yan2007,Martin, Areshkin}.
Furthermore, several works considered the reactivity of the ribbon
edges and the effects of geometrical and chemical modifications on
their electronic properties~\cite{Nakada1996, Barone, Hod, Wang,
Chang101, Gunycke, Jiang, Gowtham, Kusakabe2003, Martins,
HuangDuan, Miura93, Roman45,CanteleAPS}.
In particular, edge functionalization with oxygen-containing
groups~\cite{Hod} and edge doping or edge
imperfections~\cite{Wimmer} were shown to significantly lower
the electric field required to induce half-metallicity in
ZGNRs~\cite{Hod}, to break the spin symmetry and to promote the
possibility of net spin injection~\cite{Wimmer}. B substitutional
doping ~\cite{Martins,HuangDuan}, either on the edge or in the
bulk of GNRs, was reported to cause metal-semiconductor
transitions in the ferromagnetic and non-magnetic ZGNRs states~\cite{Martins,HuangDuanII}. The
effect of B on the AF ground state has not yet been studied.

Also of interest is the effect of edge radical functionalization
~\cite{Jiang,Kusakabe2003}. Carbon atoms at the edges of ZGNRs
offer superior chemical reactivity for the attachment of chemical
groups than those on AGNRs edges, in the ribbon's bulk, or on CNTs
edges~\cite{Jiang,Ramprasad}. Indeed, as in the case of  herringbone
and bamboo-shaped CNTs~\cite{Herringbone,Carlsson}, the reactivity
of GNRs is due to the presence of open-ended graphene sheets, which,
in the case of ZGNRs, are particularly active thanks to the presence
of electronic states highly localized at the edge carbon atoms~\cite{Jiang,Niimi}.

\begin{figure}
\includegraphics[scale=0.7]{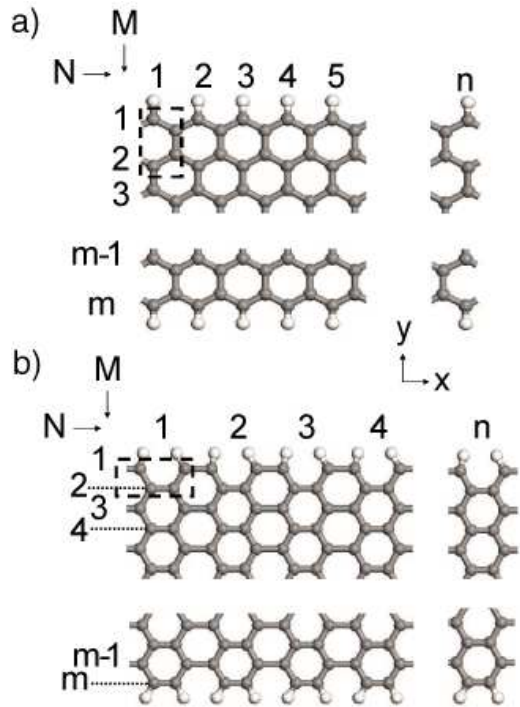}
\caption{Scheme of H terminated (a) zigzag, and (b)
armchair nanoribbons. Periodic boundary conditions are assumed in
the x direction. \emph{N} and \emph{M} are the numbers of "colums"
and "rows" of atoms used to label the \emph{M$\times$N} ribbon.}\label{Fig 1}
\end{figure}

We note that transport and Raman measurements show that as-prepared
graphene samples and devices reach the charge neutrality point for
gate voltages different from zero~\cite{Tan_Kim_condmat_2007,
Ozyilmaz_Kim_condmat_2007, MorpurgoPC, Casiraghi_condmat_2007,
DasCM, Schedin}. Most times this corresponds to a shift of the
Fermi level towards the valence band (indicative of p-doping)
~\cite{Tan_Kim_condmat_2007, MorpurgoPC, Casiraghi_condmat_2007},
but sometimes also towards the conduction band (equivalent to
n-doping)~\cite{Ozyilmaz_Kim_condmat_2007}. In particular,
refs.~\cite{MorpurgoPC,Tan_Kim_condmat_2007} seem to indicate a
bigger intrinsic p-doping for smaller ribbon size. Whether this
behavior is a consequence of chemical doping due to edge
functionalisation during lithography, or of a
rearrangement in the electronic structure due to the presence of
defect states away from the Fermi energy, or, as for large
graphene samples, due to the presence of
adsorbates~\cite{Schedin,Casiraghi_condmat_2007} needs to be
further investigated.

\section{Methodology}

We perform spin polarized \emph{ab initio} calculations with the
CASTEP plane wave DFT code \cite{Clark} on hydrogen terminated
ribbons. We use the Perdew-Burke-Ernzerhof gradient corrected
functional \cite{Perdew} and ultrasoft pseudopotentials
\cite{Vanderbilt} with cut-off energies of 400eV. Ribbons are
built from the perfect graphite geometry with an initial C-C
distance of 1.42 Å. Geometry optimization is performed for all the
structures with the Broyden-Fletcher-Goldfarb-Shanno (BFGS)
algorithm~\cite{Avriel} until all the forces are smaller than 0.04
eV/Å and the stress in the periodic direction is less than 0.05
GPa.

A 4-atom unit cell is used to build our ZGNRs as shown in
Fig.~\ref{Fig 1}(a) and the same cell rotated 90°is used to build
the AGNRs of Fig.~\ref{Fig 1}(b). The brillouin zones (BZ) of the
GNR unit cells are sampled by Monkhorst-Pack~\cite{Monkhorst}
grids of the form \emph{P$\times$}1\emph{$\times$}1 with \emph{P}
such that the maximum spacing between $k$-points in the periodic
direction \emph{x} (Fig.~\ref{Fig 1}) is < 0.1 Å$^{-1}$.
We find that, to simulate isolated ribbons, the in plane and
perpendicular distances between ribbons in adjacent super-cells
have to be larger than 5.5 Å and 6.5 Å, respectively.
The band structure is calculated with an eigenvalue tolerance of
0.001 meV.

We classify the ribbons with a \emph{M}$\times$\emph{N} convention,
where \emph{M} and \emph{N} are the number of rows and columns
across the GNR width and length, respectively, as shown in
Fig.~\ref{Fig 1}. Throughout this work, we refer to edge
functionalized ribbons with the \emph{M}$\times$\emph{N} index of
the clean GNR preceded by the chemical symbol of the
functionalizing radical and followed by the character
\emph{Z} or \emph{A} denoting the zigzag or armchair chirality.
Thus, e.g., NH$_{2}$ on the edge of a 2$\times$5 zigzag
ribbon is denoted as NH$_{2}$-2$\times$5Z.

Atom substituted ribbons are labeled by the symbol of the
substitutional atom followed by the word \emph{bulk} or
\emph{edge}, to denote bulk or edge substitution, and the
\emph{M}$\times$\emph{N} index of the ribbon. For example N bulk
substitution in the 6$\times$5 AGNR is denoted as
Nbulk-6$\times$5A.

\begin{figure*}
\centerline{\includegraphics[width=180mm]{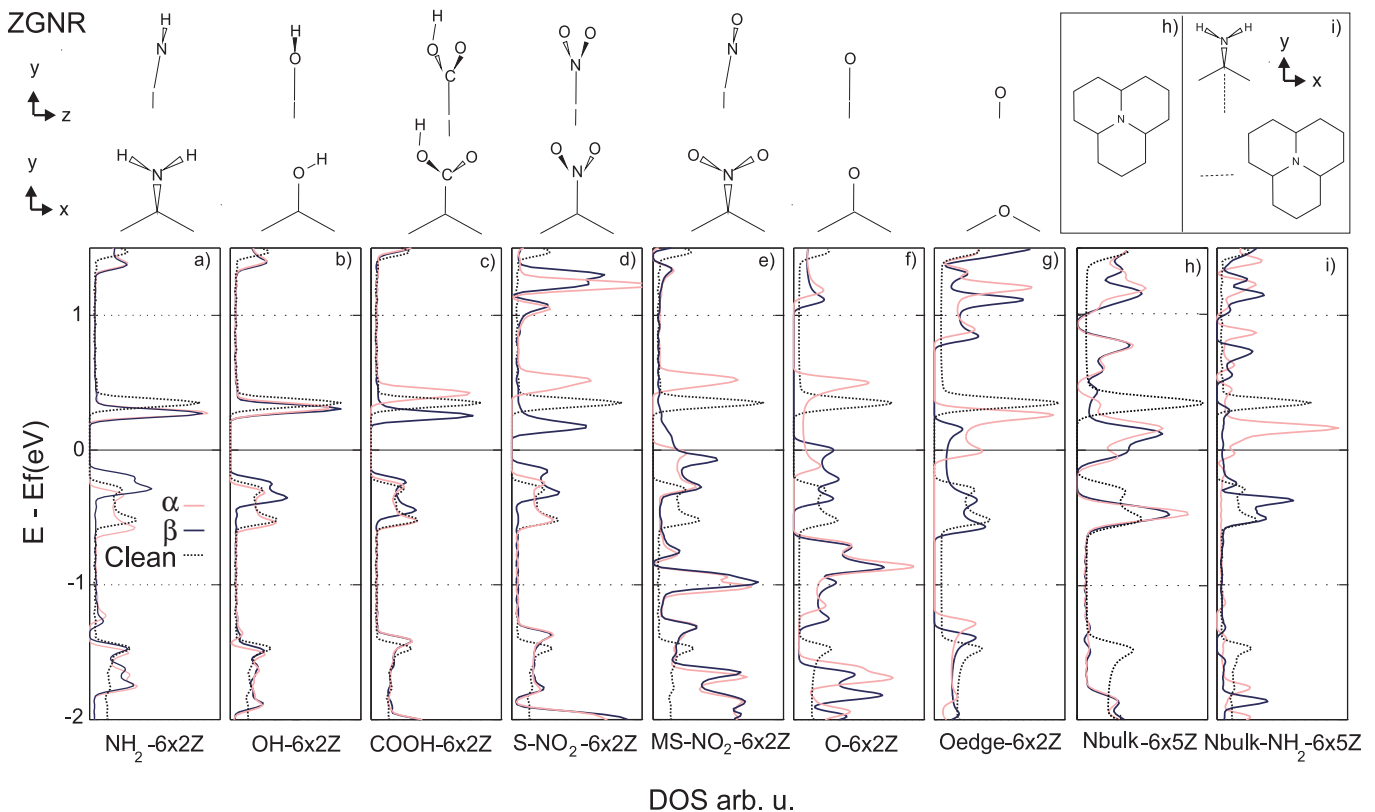}} \caption{(Color online)
Spin density of states of $\mathrm{\mathfrak{\alpha}}$ (red/gray solid
line) and $\mathrm{\mathfrak{\beta}}$ (blue/black solid line) for
functionalized ZGNRs in comparison with clean ribbon (black dotted
line). (a-f) 6\emph{$\times$}2Z (\emph{i. e.} \emph{M} = 6,
\emph{N} = 2 ZGNR) functionalized with (a) NH$_{2}$, (b) OH, (c)
COOH, (d) Stable-NO$_{2}$, (e) Meta Stable-NO$_{2}$ and (f) O
radicals in one of the ribbon edge producing breakage of the spin
degeneracy(a-f) spin-selective band gap(a-d) or
semiconductor-metal transition (e,f). (g) Oxygen edge
substitutional atom favoring a semiconductor-metal transition.
(h) 6\emph{$\times$}5 zigzag ribbon with a nitrogen bulk
substitution in the center of the ribbon. (i) Nitrogen bulk
substituted ribbon shown in (h) with an extra single NH$_{2}$
at the edge of the opposite carbon sublattice,
with the maximum distance between N and NH$_{2}$ in the periodic
direction presenting semiconductor-metal transition}. \label{Fig
2}
\end{figure*}

\section{Results}

\subsection{Edge functionalisation: single edge}
\label{subsec:singleedge}

Due to the presence of edge states in ZGNRs and their absence in
AGNRs, the effects of single-edge radical functionalization strongly
differ. Indeed, as shown in the following and in
Figs.~\ref{Fig 2}(a-f), \ref{Fig 3}(a), functionalization of ZGNRs
lifts the spin degeneracy and in some cases also promotes a semiconductor-metal
transition, while in AGNR the largest effect is the presence of impurity levels deep in the valence band that
hardly modify the gap.

We start by investigating five radical groups on
the 6$\times$2 ZGNR.
The top insets in Fig.~\ref{Fig 2}(a-f) schematize functionalisation with (a) NH$_{2},$ (b) OH, (c) COOH, (d,e)
NO$_{2}$ and (f) O. In general, geometry optimization shows that edge functionalization does not alter significantly the GNRs structure. For these radicals, we
find the following optimized geometries: the NH$_{2}$ sits 1.386
$\mathrm{\textrm{\AA}}$ away from the edge carbon, the H-N
distances are 1.021 $\mathrm{\textrm{\AA}}$ and the H-N-H plane is
tilted $\sim$20° with respect to the plane of the ribbon. The OH
remains on the graphene plane with C-O and O-H distances of 1.367
$\mathrm{\textrm{\AA}}$ and 0.977 $\mathrm{\textrm{\AA}}$
respectively and a C-O-H angle of $\sim$109°. The plane defined by
the O-C-O of the COOH has an angle of $\sim$53° with respect to
the plane of the ribbon and the C-C, C-O1, C-O2 and
O2-H distances are 1.487 $\mathrm{\textrm{\AA}}$, 1.229
$\mathrm{\textrm{\AA}}$, 1.365 $\mathrm{\textrm{\AA}}$ and 0.985
$\mathrm{\textrm{\AA}}$ respectively. For NO$_{2}$ we find two
equilibrium geometries: one energetically stable (S) and one metastable (MS),
separated by 0.407 eV. The S-NO$_{2}$ has C-N and N-O distances of
1.466 $\mathrm{\textrm{\AA}}$ and 1.257 $\mathrm{\textrm{\AA}}$, and the O-N-O plane is tilted $\sim$60° with respect
to the cross section of the ribbon. The MS-NO$_{2}$ sits in a
configuration similar to the NH$_{2}$, with C-N and N-O distances
of 1.466 $\mathrm{\textrm{\AA}}$ and 1.262 $\mathrm{\textrm{\AA}}$. Finally, for the functionalization with a single O,
this sits in the ribbon's plane in front of the edge C with C-O
distance 1.245 $\mathrm{\textrm{\AA}}$.

Functionalization strongly affects the electronic structure of
ZGNRs. This is shown in Fig.~\ref{Fig 2}(a-f), where the density
of states (DOS) of a set of functionalized ribbons is compared
with the corresponding DOS of the clean ribbon. Single-side
radical edge functionalization of a ZGNR (Figs.~\ref{Fig 2},\ref{Fig 4}) has an effect on the ribbon's spin density and
band structure similar to that of an electric field parallel to
the ribbon's plane\cite{Sonnat,Rudberg,Kan,Hod}. Thus, the presence of functional groups, while preserving an AF ground
state~\cite{SonPRL,Barone}, lifts the spin
degeneracy of the clean ribbons~\cite{Kusakabe2003}. For the
radical density we use, this results in the onset of
spin-selective semiconductivity in NH$_2$, OH, COOH and S NO$_2$
functionalizations, and of a semiconductor-metal transition
for O and MS NO$_{2}$.

The lifted spin degeneracy implies that in all cases the band gaps
of the two spin channels, which we call $\mathrm{\mathfrak{\alpha}}$-spin and
$\mathrm{\mathfrak{\beta}}$-spin, are different (Fig.~\ref{Fig
2},~\ref{Fig 4} $\mathrm{\mathfrak{\alpha}}$-red/gray solid line
and $\mathrm{\mathfrak{\beta}}$-blue/black solid line).
The change of gap with respect to the original H
terminated GNR fingerprints the radical used to functionalize the
ribbon. This is shown in Fig.~\ref{Fig 2}(a-f), which plots the DOS
for each of the five radicals.
A pristine 6$\times$2Z ribbon has a 0.58 eV direct band gap at
\emph{k} = 0.5$\pi/c$, \emph{c} being the size of the ribbon's
unit cell along its periodic direction. The different
functionalizations change the band gap as follows. In presence of
NH$_{2}$ the gap remains direct and at \emph{k} = 0.5$\pi/c$ for
both spins, but shrinks to 0.42 eV and 0.39 eV for the
$\mathrm{\mathfrak{\alpha}}$ and $\mathrm{\mathfrak{\beta}}$ spin
channels, respectively. A similar effect is obtained by means of OH
functionalisation, with final gaps of 0.57 eV and 0.49 eV. In the
case of COOH, the direct gap at \emph{k} = 0.5$\pi/c$ increases up
to 0.62 eV, but an indirect one of 0.45 eV,  with extremes at
\emph{k} = 0 in the conduction band and \emph{k} = 0.5$\pi/c$ in
the valence band, appears. Similarly, S-NO$_{2}$ gives a direct
gap of 0.66 eV and a smaller indirect gap of 0.28 eV with the
extremes located at the same position as the indirect gap due to
COOH. MS-NO$_{2}$ and O have several extra levels in the
valence band and a zero gap for both spin channels. This makes
them metallic.

The semiconductor-metal transition we predict for O functionalized
ZGNRs is an intriguing. Ribbons are often produced
by oxygen plasma etching~\cite{Ozyilmaz_Kim_condmat_2007},
suggesting that the same technique used to cut ribbons, could also
influence their electronic properties. We thus perform a calculation for a lower edge functionalization density. We find that for a
6$\times$6Z ribbon the semiconductor-metal transition does not
happen, but we have a gap of 0.82
eV and 0.41 eV for the $\mathrm{\mathfrak{\alpha}}$ and
$\mathrm{\mathfrak{\beta}}$ spins, respectively (+0.40\% and -30\%
with respect to pristine ribbon). This implies that
the semiconductor/metallic behavior of the ribbon could be controlled by
the degree of oxidation of a single zigzag edge, provided this does not revert to OH, which, as discussed before, would not significantly affect the gap.

We note that, while NO$_{2}$ and NH$_{2}$ on graphene act as
strong acceptor and donors~\cite{Schedin, Wehling}, the same
molecules at the edges of the ZGNRs modify the
electronic structure in the gap vicinity, but do not act as dopants.

\begin{figure}
\includegraphics[scale=0.95]{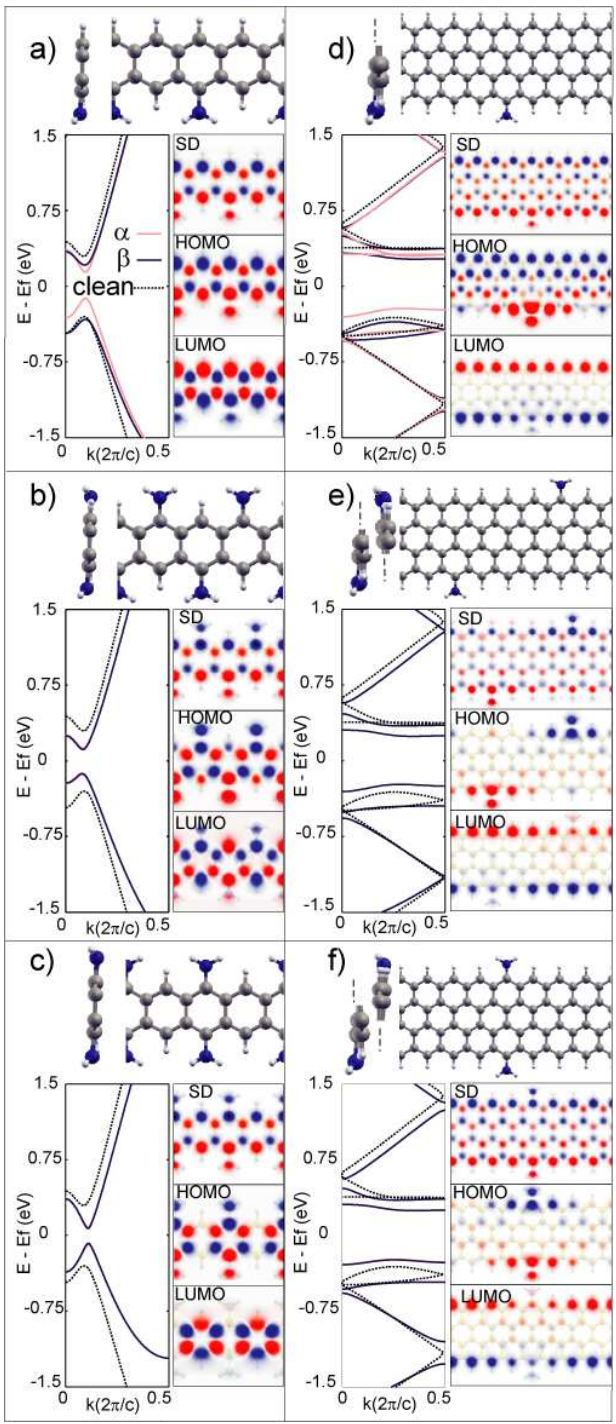}
\caption{(Color online) Band structure of the NH$_{2}$ 2\emph{$\times$}2Z
(a-c) and 4\emph{$\times$}8Z (d-f) single-side (a,d) and
double-side (b,c,e,f) functionalization. Single side
functionalization (a,d) breaks spin degeneracy,
whereas this is nearly recovered for double-sided
antisymmetric (b,e) and symmetric (c,f) functionalization. $\mathrm{\mathfrak{\alpha}}$($\mathrm{\mathfrak{\beta}}$)
spins are represented by red/gray(blue/black) solid lines. The
folded band structures of the pristine ZGNRs are plotted with black
dotted lines. On the right of each BS are the corresponding spin
density maps ($\mathrm{\mathfrak{\beta}}$-blue and
$\mathrm{\mathfrak{\alpha}}$-red) of the total spin density (SD),
spin density of the HOMO and LUMO bands. The
image shows the relaxed structure in \emph{xy} projection and a
close up of the edges in the \emph{yz} projection.} \label{Fig 5}
\end{figure}

Fig.~\ref{Fig 5} shows the effect of functionalization on the spin
density and band structure of both single- and double-side
functionalized ZGNRs. Black dotted lines represent the bands of the 2\emph{$\times$}1(4\emph{$\times$}1) clean ZGNRs
folded into the 2\emph{$\times$}2(4\emph{$\times$}8) reciprocal
unit cell. The band structure of the clean ribbons is spin
degenerated~\cite{SonPRL,Barone}, and the spin density is
distributed in an AF arrangement from the edges to the center, as in Refs.\cite{SonPRL,Barone}.
Fig~\ref{Fig 5}(a,d) indicates that single-edge NH$_{2}$
functionalization introduces extra charges that alter the spin
density.
Due to the zigzag geometry, C atoms sitting on the outmost
position of opposite edges always belong to different sublattices.
As a consequence, single edge functionalization affects
prevalently the spin channel localized on the same edge where the
functional group is attached.
This is shown in Fig.~\ref{Fig 5}(a,d). Indeed, after
functionalization, the two sub-lattices are visible in the total spin density (SD) maps with the
characteristic AF spin arrangement. However,
NH$_{2}$ alters the SD, attracting
$\alpha$ spin regions in its vicinity, while the $\beta$ spin
density is substantially unchanged.

The perturbation of the SD is reflected in the lifting of the
spin degeneracy illustrated in the DOS plot of Fig.~\ref{Fig 2}, and
in the band structure of Fig.~\ref{Fig 5}(a,d).
As the highest occupied molecular orbital (HOMO) and lowest
unoccupied molecular orbital (LUMO) are highly localized on the
edge of the ZGNRs~\cite{Nakada1996,SonPRL}, the influence of edge
functionalization is predominantly on these orbitals. This is
shown by the maps of the total spin density and of the HOMO and
LUMO spin denisity in Fig.~\ref{Fig 5}(a,d).
For the valence band orbitals of the 2\emph{$\times$}2 and
4\emph{$\times$}8 ribbons, the $\mathrm{\mathfrak{\alpha}}$ spin
(red/gray) hosting the NH$_{2}$ presents a stronger
modification than the $\mathrm{\mathfrak{\beta}}$ spin
(blue/black) orbital in the opposite edge.
Looking to the corresponding band, the
$\mathrm{\mathfrak{\alpha}}$ HOMO is more modified than the
$\mathrm{\mathfrak{\beta}}$ HOMO with respect to the original HOMO
(black dotted line).
The breakage of the spin degeneracy at the top of the HOMO band is
0.21 eV and 0.13 eV for the NH$_{2}$-2\emph{$\times$}2Z and
NH$_{2}$-4\emph{$\times$}8Z ribbons respectively (note the color
matching between spin density maps and spin bands in all
Fig.~\ref{Fig 5}).
We expect similar effects for any \emph{N}, \emph{M}
combination and for any radical attached to only one ribbon edge.

We finally consider edge radical functionalization on AGNRs. AGNRs
do not have edge states near the band gap, so the effect is expected to be weaker than in ZGNRs.
Fig.~\ref{Fig 3}(a) shows NH$_{2}$ functionalization on the edge
of the 12\emph{$\times$}3 AGNR. Here an impurity level is found $\sim$1.5 eV below the edge of the valence
band, the gap increases by +0.02 eV ($\sim$4\%), with no lifting
of the spin degeneracy. Thus, for AGNR, NH$_{2}$ edge
functionalization does not significantly affect the electronic
properties, since the impurity level is located deep in the
valence band and no other major change happens to the DOS. We
expect that functionalization with other chemical groups will give
similar results. Nevertheless, finding a chemical group able to
introduce impurity levels in the gap will be of great interest for
electronic applications, so further investigations are desirable.

\begin{figure*}
\includegraphics[scale=0.9]{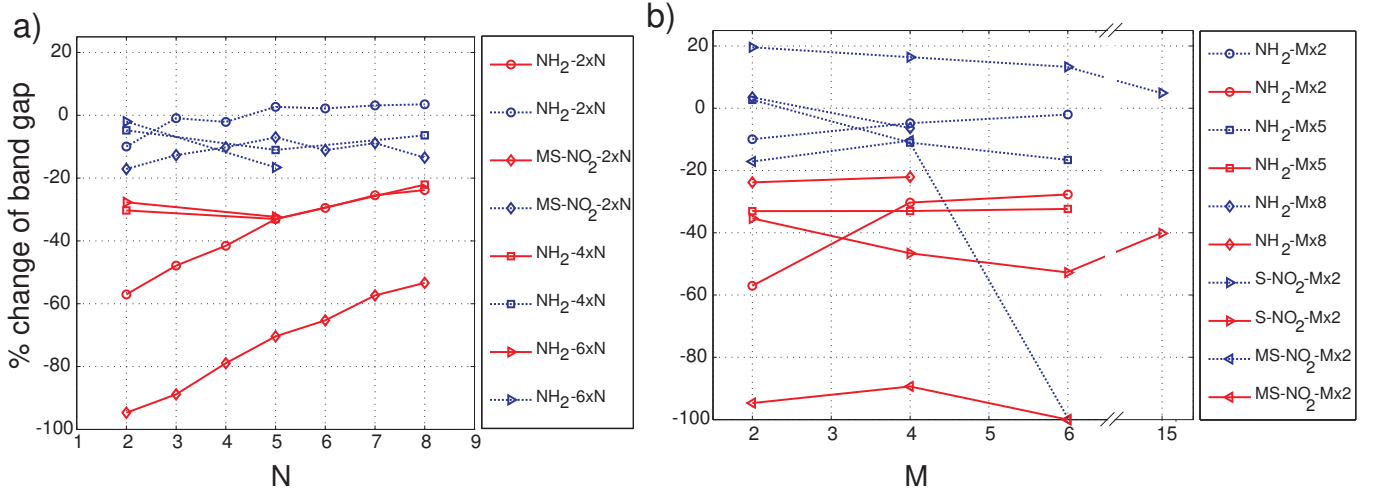}
\caption{(Color online)Change of spin band gap ($\mathrm{\mathfrak{\beta}}$
channel-blue dotted line, $\mathrm{\mathfrak{\alpha}}$ channel-red
solid line) as a function of (a) edge density \emph{-N-} and (b)
width of ribbon \emph{-M-} for different set of edge
functionalized ZGNRs.\label{Fig 4}}
\end{figure*}

\subsection{Edge functionalisation: double edge}
\label{subsec:doubleedge}

The effects of double edge functionalization are shown in
Fig.~\ref{Fig 5}(b,c,e,f). Interestingly, when another radical of
the same species is located in the opposite edge of the ribbon,
the spin degeneracy is nearly restored . Fig.~\ref{Fig 5}(b,e)
show the 4\emph{$\times$}8Z and 2\emph{$\times$}2Z ribbons with
two NH$_{2}$ impurities on opposites edges in an asymmetric
arrangement. The breakage of the degeneracy between
$\mathrm{\mathfrak{\alpha}}$ and $\mathrm{\mathfrak{\beta}}$ HOMO
is only 1 meV and 0.4 meV for the double side asymmetric
functionalized 2\emph{$\times$}2Z and 4\emph{$\times$}8Z ribbons,
respectively. Fig.~\ref{Fig 5}(c,f) show the same ribbons with two
NH$_{2}$ impurities sitting on opposite edges in a symmetric arrangement.
In this case the breakage of the spin degeneracy between
$\mathrm{\mathfrak{\alpha}}$ and $\mathrm{\mathfrak{\beta}}$
HOMO bands is 11 meV and 3.6 meV for the 2\emph{$\times$}2Z and
4\emph{$\times$}8Z ribbons respectively.

The differences between the symmetric and antisymmetric geometries
indicate the different stability of these configurations. Indeed,
the total energy of the antisymmetric ribbon is lower than that of
the symmetric ribbon by 42 meV and 34 meV for the
2\emph{$\times$}2Z and 4\emph{$\times$}8Z ribbons.

To restore spin degeneracy in 2-sided functionalized
ribbons, it is necessary to have the same impurity density on
both edges, and the maximum separation between consecutive
radicals of opposite edges has to be achieved. Finally, it is
important to stress that any geometry distortion or irregularity
will result in a stronger magnetization of one of the edges, thus lifting the spin degeneracy.

\subsection{Edge functionalisation: density of radicals dependence}
\label{subsec:density}

We now consider the effects of the spacing between radicals, which
is proportional to the index $N$, on the electronic structure of
one-sided functionalized ZGNRs. Fig.~\ref{Fig 4}(a) plots, for a
set of NH$_{2}$ and NO$_{2}$ functionalized ribbons, the change in
the band gap for each spin as a function of $N$. A high level of
NH$_{2}$ functionalization of the $M=2$ ribbon can shrink the band
gap of the $\alpha$ spin up to $\sim$60\%, while leaving the $\mathrm{\mathfrak{\beta}}$
gap almost unchanged.

A similar behavior is observed for the
MS-NO$_{2}$-2\emph{$\times$}$N$Z ribbon. In this case, increasing
the functionalization densities narrows the $\alpha$ gap up to
almost its complete closure, while the $\beta$ gap has a maximum
change of $\sim$-20\%, almost independent of doping. These
results, obtained for very thin ribbons ($M=2$), seem to suggest a
correlation between the breakage of the spin degeneracy and the
edge-functionalization density, the former slowly decreasing for
increasing $N$. However, in wider ribbons \emph{(e.g.} $M$ = 4,
6), the band gap shows a much weaker dependence on $N$. In
particular, for the NH$_{2}$-4$\times$$N$ ribbon the changes in
the $\mathrm{\mathfrak{\alpha}}$ and $\mathrm{\mathfrak{\beta}}$
gaps oscillate between 20-35\% and 3-18\% respectively when $N$
spans between 2 and 8. Almost identical gap variations are
observed for NH$_{2}$-6$\times$$N$Z for $N$ between 2 and 5.

\begin{figure*}
\centerline{\includegraphics[width=180mm]{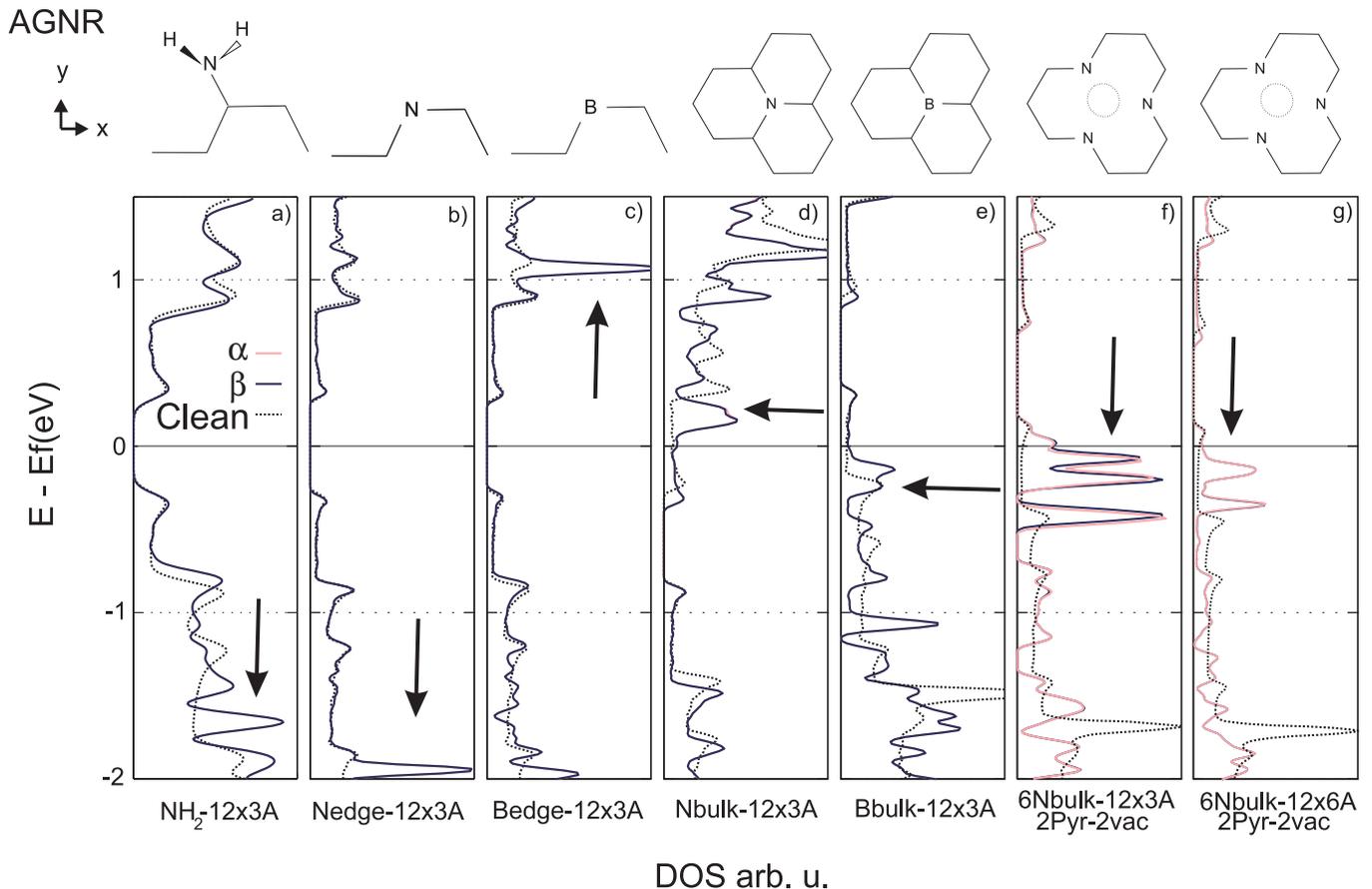}} \caption{(Color online)
Spin density of states of $\mathrm{\mathfrak{\alpha}}$ (red/gray solid
line) and $\mathrm{\mathfrak{\beta}}$ (blue/black solid line) for
functionalized AGNRs in comparison with clean ribbon (black dotted
line). (a) 12$\times$3 AGNR functionalized in one edge with
NH$_{2}$. (b) N, (c) B edge substitutional atoms on
the 12x3 AGNR. N and B bulk substitution for the 12x3 AGNR is shown
in (d) and (e), where a semiconductor-metal transition occurs.
(f,g) two Pyridine-like substitutional doping in the bulk of 2x3a and 12x6a
presenting metal behavior. Arrows indicate impurity levels due to
substitutional atoms. The DOS of the clean ribbon is shifted for
the bulk N and B substitution such that the edge of the valence
band of clean and doped ribbons coincide}\label{Fig 3}
\end{figure*}

\subsection{Edge functionalisation: width dependence}
\label{subsec:width}

Fig.~\ref{Fig 4}(b) plots the variation of the $\alpha$ and
$\beta$ bandgap, relative to the size of the bangap in the clean
ribbon, in single-side functionalized ribbons as a function of
$M$. Fig.~\ref{Fig 4} suggest that the modulation of the gap due
to functionalization depends very on $M$, although some
strong fluctuations may still occur for the thinnest ribbons
(\emph{i.e. M} = 2).
The points for~\emph{M} = 15 in Fig.~\ref{Fig 4}(b) confirm
this behavior, showing that the variation of the $\alpha$-spin gap
due to functionalization in the S-NO$_{2}$-$M$$\times$2Z ribbons is -32\% to -55\% for $M$ changing from 2 to 6. A further
increase of $M$ from 6 to 15 (ribbon width increasing from 1.3 nm
to 3.8 nm) leaves the functionalization contribution almost
unchanged for both spin channels.
Fig.~\ref{Fig 4} thus confirms that edge functionalization always
promotes lifting of the spin degeneracy, and indicates that such effect decays slowly as a function of
\emph{M}.

A remarkable exception are the MS
NO$_2$-ZGNRs.  The MS-NO$_{2}$-2$\times$2Z
and MS-NO$_{2}-4\times$2Z are half semiconducting, and
the $\alpha$ and $\beta$ gaps are respectively reduced to
$\sim$90\% $\sim$15\% of the original gap, while the
MS-NO$_{2}$-6$\times$2Z ribbon is metallic for both spin channels.

\subsection{Atomic substitution}
\label{subsec:atomicsubs}

In semiconducting CNTs, substitutional B~\cite{Yi,Carrol1998}
and N~\cite{Nevidomskyy2003,Yi} act as acceptors or donors,
inserting levels in the gap or causing a semiconductor-metal
transition. B substitution on the edge of meta-stable ferromagnetic ZGNRs was shown to induce metal-semiconductor transitions
\cite{Martins}. It is thus interesting to study in more detail the
effects of atomic substitutions in GNRs.

We start by considering O substitution on the edge of ZGNRs. By
using a 6$\times$2Z ribbon, we find that in the optimized geometry
the two C-O distances are 1.41\AA. The resulting DOS is reported in Fig.~\ref{Fig
2}(g), showing that the appearance of new states in the
conduction band promotes the onset of a semiconductor-metal
transition.

To investigate the role of the functionalization density, we
repeat the same calculation for a 6$\times$6Z ribbon. In this case,
no semiconductor-metal transition is observed. Instead, the spin
degeneracy is lifted, and the band gap for the two spin channels
are 0.449eV and 0.399eV, which represent a change of
-22\% and -31\% with respect to the pristine ribbon.
Overall, the effects of O functionalization (Fig.~\ref{Fig 2}(f))
and edge substitution (Fig.~\ref{Fig 2}(g)) differ only for the
the position of the extra levels.
Since in the center of the ribbon the spin density is much lower
than the edges, the effect of bulk substitutions on the
breakage of the spin degeneracy is expected to be weaker than the
corresponding edge substitution or edge functionalization. Fig.~\ref{Fig 2}(h) shows the effect of N substitution in the center of a $6\times5$Z ribbon, which corresponds 1.7 at\%, \emph{i.e.} 1 N atom in a lattice of
59 C atoms. This gives a
semiconductor-metal transition, with a weak spin degeneracy lifting.
However, as shown in Fig.~\ref{Fig 2}(i), the introduction on the
edge of the same ribbon of a NH$_{2}$ radical leads to a much
more pronounced difference between the two spin
channels DOS. In this case, the N substitutional atom is placed in the
opposite carbon sub-lattice with respect to NH$_{2}$, and the
separation between N and the NH$_{2}$ radical is maximized. This
arrangement is chosen with the aim to alter the AF
state of both sub-lattices.

We now consider low concentration
atomic substitution on the edge and bulk of AGNRs. For
edge substitution we use the
12\emph{$\times$}3 AGNR (1 substitutional atom in 71 C atoms, 1.4
at\%). In this case N impurity states appear $\sim$1.6 eV
below the valence band (Fig.~\ref{Fig 3}(b)) and the
B states $\sim$0.8 eV above the conduction band
(Fig.~\ref{Fig 3}(c)). The impurities modify the
gap by +8.7 meV (+1.5\%) for N and -32 meV (-5\%) for B. In the
relaxed structure the two N-C nearest neighbor distances are 1.33
Å and 1.36 Å and the B-C bonds are 1.40 Å and 1.43 Å.

Bulk substitution is then investigated in the 12\emph{$\times$}3 AGNRs. Here
we observe that both N and B give impurity levels localized around
the substituted atoms. This shifts the Fermi energy $\sim$0.5 eV
inside the former conduction and valence band,
resulting in a semiconductor-metal transition. As indicated by the
arrows in Fig.~\ref{Fig 3}(d,e), the impurity bands are localized
at $\sim$ 0.15 eV from the Fermi energy. In the relaxed structure
the N-C and B-C distances are 1.41 Å and 1.49 Å,
respectively.

Unlike CNTs~\cite{Carrol1998,Yi,Nevidomskyy2003}, substitutional B and N in AGNRs do not give
electronic states in the band gap, even if semiconductor-metal
transitions can still be observed.
Further investigations with different substitutional atoms or
geometries are needed to determine if and how conventional doping
of GNRs can be achieved.

We finally consider the effect of pyridine-like doping. This
consists in the substitution of 4 carbon atoms by 3 nitrogens and
a vacancy, as schematically shown in the diagram of Fig.~\ref{Fig
3}(f,g). In CNTs, high density pyridine-like doping was found to
introduce donor-like states above the Fermi energy and to produce
semiconductor-metallic transitions~\cite{Czerw}. We start by
introducing 2 pyridine-like impurities in the 12\emph{$\times$}3
AGNR, corresponding to 9.4 at\% N. The
vacancies are arbitrarily positioned near the center of the ribbon
at 1 row and 2 columns of atoms away from each other. This gives 4 localized levels in the valence band near
the Fermi energy, which induce a Fermi energy shift and cause a
semiconductor-metal transition. We also consider two
pyridine-like impurities in the 12\emph{$\times$}6A
ribbon, corresponding to 4.4 at\% N. These are
positioned in the nearest possible sites to the center of the
ribbon at a inter-vacancy distance in the periodic direction of 5
columns of carbon atoms. The arrow in Fig.~\ref{Fig 3}(g) shows
the impurity levels in the valence band of the ribbon due to the
pyridine-like doping. These levels, as in the case of higher
impurity density, shift the Fermi energy and cause a
semiconductor-metal transition.

\section{Discussion }

The fabrication of devices with a new material
requires the understanding all the factors that influence its
electronic and spin properties.
The geometry and the presence
of various form of chemical modifications are crucial for GNRs. The geometry is defined by length, width, and edges chirality, whereas chemical modifications can
be caused by different edge terminations, and
by substitutional atoms in the ribbons' body.

In this paper, we examined how all these factors affect
the electronic and spin properties of infinite ribbons. Our main result is
that a strong spin polarization is maintained in all the simulated
zigzag ribbons. In particular, our calculations show that single
edge functionalization leads to half-semiconductors with
different band gap for each spin and can also result in 
a spin-polarized half-semiconductor or in a semiconductor-metal transition.
This implies the possibility of using single edge functionalized
ribbons to realize spin filtering devices, even though single-edge functionalisation may be experimentally very challenging. On the other hand, the weak dependence of spin degeneracy lifting and half-metallicity on width is encouraging for potential applications, since wider ribbons are easier to make.

We also observe that high concentrations of oxygens on
the ribbons edges are likely to produce metallic ZGNRs, while for
lower oxygen concentrations, ZGNRs are expected to be half semiconducting. Thus, oxygen is one of the most
effective elements to promote the closure of the electronic gap
for both spin channels. Since ribbons can be cut from bulk
graphene by means of an oxygen plasma~\cite{Ozyilmaz_Kim_condmat_2007,Han}, this could affect their final
electronic properties. However, it is likely that OH may also result from this treatment. But OH does not close the gap. Given the large gaps so far reported~\cite{Ozyilmaz_Kim_condmat_2007,Han}, it is thus unlikely that these GNRs are just O terminated.

It is known that the electronic structure of AGNRs does not
present spin-polarization effects~\cite{SonPRL,Nakada1996} and that their gap scales inversely
with the width~\cite{SonPRL,Nakada1996}.
Our calculations indicate that the gap in AGNRs is extremely
stable with respect to the edges chemical termination. This is due to the absence of edge states, which
is, on the contrary, the main characteristic of zigzag ribbons.
Armchair ribbons seem thus ideal for devices.
However, our calculations indicate
that the presence of B and N does not give
defect states that pin the Fermi energy in the gap, showing
that the behavior of AGNRs with respect to atomic substitutions is
markedly different from that of CNTs.
Indeed, even if it is well established that replacing C atoms with
B (N) in semiconducting CNTs results in electronic states that pin the Fermi energy close to the
conduction (valence) band, producing a
$n(p)-$doping~\cite{HirschFunc, Georgakilas,
Carrol1998,Yi,Nevidomskyy2003,Czerw}, our simulations show that in
AGNRs the same substitutions lead to completely different effects.
If the substitution occurs on the ribbon edges, the impurity
levels are always very far from the Fermi energy, and the
electronic properties of the ribbon are substantially unchanged.
If it occurs in the center, the defect states appear
close to the bottom (top) of the conduction (valence) band of the
pristine ribbon, but the Fermi energy is shifted in the
conduction (valence) band, triggering a semiconductor-metal
transition, rather than doping, unlike in CNTs and bulk semiconductors.

The physical reason for the different effects of chemical doping in
GNRs and CNTs is likely to derive from the different boundary
conditions to which the electronic wave-functions have to obey in
the two materials. However, further investigations are needed to
fully address this point. The reason why edge
substitutional atoms do not result in a shift of the Fermi
energy is related to the relative position of defect
states and the bandgap of the pristine ribbon.
Indeed, N edge atoms are attached to the ribbon with very
localized chemical bonds in an $sp^3$ configuration. This promotes
the formation of a deep state in the valence band, which is
occupied by the extra electron carried by the N.
On the contrary, N in the bulk of the ribbon is in an
$sp^2$ configuration. This gives delocalized
states in the conduction band. Since no new states
are formed in the valence band, the extra electrons coming from N
have to be allocated in what was the conduction band of the
pristine ribbon, with a consequent shift of the Fermi energy.
A perfectly specular consideration can be done for B doping.

Finally, we stress that here we focussed only on chemical modification of the
ribbons, without taking into account additional effects, such as chemisorption or
physisorption of molecules on the ribbons'surfaces.
Since adsorbates can easily shift the fermi level in large graphene samples
~\cite{Schedin,Wehling,Leenaerts,Uchoa,Zanella}, the same could also happen for ribbons.
This deserves further investigation, and may help understand
some experimental observations, such as the tendency of small ribbons to
appear positively charged~\cite{Tan_Kim_condmat_2007, MorpurgoPC, Casiraghi_condmat_2007}
or the dependence of their minimum conductivity on the environment~\cite{Chen}.

\section{Conclusions}

We studied atomic substitution and edge functionalization in graphene ribbons. The
distinctive anti-ferromagnetic phase of the spin on the edges of
zigzag nanoribbons is altered by any single edge
functionalization, producing half-semiconducting structures. An
appropriate functionalization with the correct radical could produce half-metallicity, since different radicals alter
the anti-ferromagnetic arrangement in different ways. Oxygen edge
substitution and functionalization at high density in zigzag
nanoribbons produce semiconductor-metal transitions. Double edge
functionalization of zigzag ribbons can reduce the semiconducting gap, with an antisymmetric
arrangement of the radicals energetically favored. Nitrogen bulk substitution at 1.7
at\% in zigzag ribbons also makes them metallic. The gap of armchair ribbons is robust
against edge functionalization, and N,B edge substitutions. However N, B and pyridine-like bulk substitution causes
semiconductor-metal transitions.

\section{Acknowledgments}

F.C.S. acknowledges funding from CONACYT Mexico. S.P. acknowledges
funding from The Maudslay Society and Pembroke College, and A.C.F.
from The Royal Society and The Leverhulme Trust.

\end{document}